\documentclass[conference]{IEEEtran}
\usepackage[export]{adjustbox}
\usepackage{soul}
\usepackage{colortbl}
\usepackage{xcolor}
\usepackage{tabularx}
\usepackage{hhline}
\usepackage{pgfplots}
\usepackage{pgfplotstable}
\pgfplotsset{compat=1.7}
\usetikzlibrary{dsp,chains,arrows,calc,positioning}
\usetikzlibrary{shapes.multipart, decorations.pathreplacing}
\usepackage{subfigure}
\usepackage{mathtools}

\def\nbd{{\mathbf{d}}}

\def\nbs{{\mathbf{s}}}

\def\nbu{{\mathbf{u}}}

\def\nbw{{\mathbf{w}}}
\def\nbx{{\mathbf{x}}}

\def\nb0{{\mathbf{0}}}
\def\nb1{{\mathbf{1}}}

\def\nbD{{\mathbf{D}}}

\def\nbF{{\mathbf{F}}}

\def\CN{\mathcal{CN}} 

\def\ncalU{{\mathcal{U}}}

\def\E{\mathbb{E}}

\def\Exp{\mathbb{E}}

\def\C{{\mathbb{C}}}

\newcommand{\herm}{^{\text{\sf H}}}

\usepackage{tikz}
\usepackage{circuitikz}
\usetikzlibrary{dsp,chains,arrows,calc,positioning}
\usetikzlibrary{shapes.multipart, decorations.pathreplacing}

\usetikzlibrary{arrows}
\usepackage{cite}
\usepackage{amsmath,amssymb,amsfonts}
\usepackage{algorithmic}
\usepackage{graphicx}
\usepackage{textcomp}
\usepackage{siunitx}
\usepackage{multirow}
\def\BibTeX{{\rm B\kern-.05em{\sc i\kern-.025em b}\kern-.08em
    T\kern-.1667em\lower.7ex\hbox{E}\kern-.125emX}}

\definecolor{lightGreen}{rgb}{0.5313, 0.7875, 0.7877}
\definecolor{mediumGreen}{rgb}{0.3296, 0.6354, 0.6358}
\definecolor{darkGreen}{rgb}{0.1154, 0.4519, 0.4522}

\definecolor{babyblueeyes}{rgb}{0.63, 0.79, 0.95}

\usetikzlibrary{decorations.markings}
\input{lib/basestation}

\begin{document}

\title{Understanding Energy Efficiency and 
Interference Tolerance 
in Millimeter Wave Receivers}

\author{
    \IEEEauthorblockN{Panagiotis 
    Skrimponis\IEEEauthorrefmark{1}, 
    Seongjoon Kang\IEEEauthorrefmark{1},
    Abbas Khalili\IEEEauthorrefmark{1},
    Wonho Lee\IEEEauthorrefmark{2},
    Navid Hosseinzadeh\IEEEauthorrefmark{2},
   }
    
    \IEEEauthorblockN{
    Marco Mezzavilla\IEEEauthorrefmark{1},
    Elza Erkip\IEEEauthorrefmark{1},
    Mark J. W. Rodwell\IEEEauthorrefmark{2}, 
    James F. Buckwalter\IEEEauthorrefmark{2},
    Sundeep Rangan\IEEEauthorrefmark{1}}
    
    \IEEEauthorblockN{
    \IEEEauthorrefmark{1}NYU Tandon School of Engineering,
    New York University, Brooklyn, NY}
    \IEEEauthorblockN{
    \IEEEauthorrefmark{2}Dept. ECE,  University of
    California, Santa Barbara}
    
    \thanks{P. Skrimponis, S. Kang, A. Khalili, M. Mezzavilla, E. Erkip and S. Rangan are with NYU Wireless, Tandon School of Engineering, New York University, Brooklyn, NY.
    They are supported under
    NSF grants 1952180, 1925079, 1564142, 
    1547332, the Semiconductor Research Corporation (SRC) and the industrial affiliates
    of NYU Wireless.
    }

}
\IEEEoverridecommandlockouts

\maketitle

\begin{abstract}
Power consumption is a key challenge
in millimeter wave (mmWave) receiver front-ends,
due to the need to support high dimensional antenna arrays
at wide bandwidths.  
Recently, there has been considerable work in developing 
low-power front-ends, often based on low-resolution ADCs
and low-power mixers.
A critical but less studied consequence of such designs is the
relatively low-dynamic range which in turn exposes the receiver 
to adjacent carrier interference and blockers.
This paper provides a general mathematical framework for 
analyzing the performance of mmWave front-ends
in the presence of out-of-band interference.
The goal is to elucidate the
 fundamental trade-off of power consumption, interference
tolerance and in-band performance.  
The analysis is combined with detailed network simulations
in cellular systems with multiple carriers, as well as detailed circuit
simulations of key components at \SI{140}{GHz}.  The analysis
reveals critical bottlenecks for low-power interference robustness
and suggests designs enhancements  for use in
practical systems.
\end{abstract}

\section{Introduction} \label{sec:intro}
A key challenge in millimeter wave (mmWave)
receiver front-ends is power consumption, particularly
for mobile and portable devices.
High power consumption is especially challenging
in emerging systems above \SI{100}{GHz} that need to support
a large number of array elements at high bandwidths
with relatively poor device efficiency \cite{mendez2015channel,rappaport2019wireless,skrimponis2020efficient,skrimponis2020power,yan2020dynamic,tan2020thz}.

Significant recent progress has been made with 
reduced power
architectures, most notably via low-resolution analog-to-digital converters (ADCs)
and phase shifters, as well low-power mixers \cite{naqvi2018review,abbas2017millimeter,zhang2018low,abdelghany2018towards,yan2019linearization,mo2017hybrid,jacobsson2017throughput}.  
A common theme in these designs
is to sacrifice dynamic range for lower power.
This design choice is often based on the fact that
communication systems in the mmWave bands typically operate
at relatively low signal-to-noise ratios (SNRs) per antenna
that can be recovered from beamforming.

However, most prior analyses of low-dynamic range
architectures have generally only considered the 
\emph{in-band} signal distortion.
Aggressive use of low dynamic range front-ends
introduces potential susceptibility of the receiver
to high power \emph{out-of-band} or 
adjacent carrier signals~\cite{marti2021hybrid,marti2021jammer,akhlaghpasand2020jamming}. 
These adjacent carrier signals can be particularly
large in uncoordinated 
cellular deployments with multiple carriers
as well as unlicensed use
\cite{rebato2017hybrid}.

\begin{figure}[t]
    \centering
    \includegraphics[width = 0.99\linewidth]{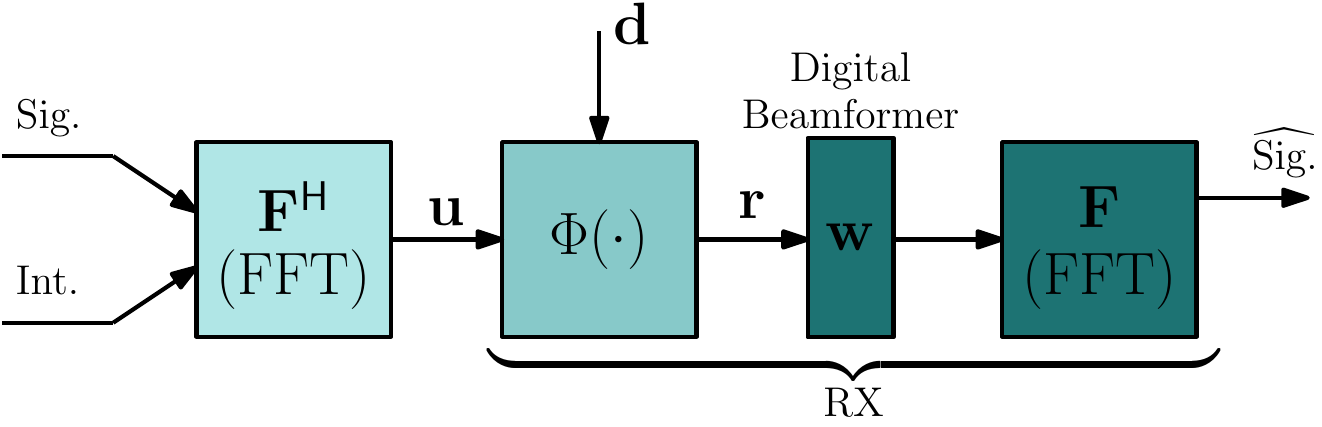}
    \caption{Abstract mathematical model for a system
    with adjacent carrier interference and non-linear,
    noisy front-end. Sig. and Int. denote the desired signal and interference signal from the other transmitters, respectively.}
    \label{fig:abstract_model}
\end{figure}

For receiver design in conventional bands below \SI{6}{GHz},
it is well-known that interference rejection 
(also called \emph{blockers}) 
can be the dominant
driver of power consumption.
Some example low-power designs with good
out-of-band
blocker performance can be found in
\cite{sepidband2018cmos,trotskovsky20180,agrawal2018interferer}.
However, the design and analysis of 
low power designs under adjacent carrier
interference in the mmWave
range face unique challenges. 
Most importantly, standard
SAW and BAW (surface acoustic wave and bulk
acoustic wave) filters are not easily available in
these frequencies~\cite{mahon20175g}.
Recent work has attempted to use filters integrated into the package
\cite{watanabe2020review,siddiqui2021dual,gu2021antenna}, but these come
with added RF signal loss.  In addition, the nonlinearities inherent
in low-power mmWave designs can be difficult
to analyze and mitigate against.
For example, there has been some recent work
studying jamming signals on 
mmWave receivers with low-resolution ADCs
and other non-linear hardware impairments in
\cite{akhlaghpasand2020jamming,pirzadeh2019mitigation,marti2021hybrid,jacobsson2018massive}.
However, these works generally consider the case
where the jamming signal is in-band,
but spatially separated 
from the desired signal.  In this work,
we focus on adjacent carrier interference
where interfering signal is in a different
frequency band, as would occur in cellular,
licensed deployments.

For mmWave systems under adjacent
carrier interference, we thus wish to understand
the fundamental relation between power 
consumption and interference tolerance
and how this trade-off can be best optimized.
Towards this end, 
the contributions of the paper are three-fold:
First,  building on the work in
\cite{dutta2020capacity,skrimponis2020power,skrimponis2021towards,skrimponis2020efficient},
we provide a general methodology 
for mathematically analyzing the sensitivity of
receivers from adjacent carrier emissions.
Importantly, the framework can incorporate 
general power spectral densities of the interfering
signals, models
of the filters at various points in the receiver
chain, as well as non-linearity and quantization 
limits in the analog front-end.

Secondly, we apply the methodology to practical
designs
of \SI{28}{GHz} and \SI{140}{GHz} receivers.  
The analysis uses detailed circuit
and signal processing simulations to provide
realistic estimates on the power consumption
and elucidate the main bottlenecks in interference rejection
robustness.

Finally, we perform simple network simulations
at \SI{28}{GHz} and \SI{140}{GHz}
to estimate the frequency of high-power adjacent carrier
signals in likely cellular deployments with multiple carriers.
The simulations consider cases with non-co-located
cell sites,
which introduce the highest level of interference
\cite{rebato2017hybrid}.


\section{System Model} \label{sec:model}

The model is an extension of the analysis framework in \cite{dutta2020capacity,skrimponis2020power,skrimponis2021towards,skrimponis2020efficient},
which studied a single transmitter and receiver. Here, we extend
the model to add interfering transmitters.
The basic set-up is shown in Fig.~\ref{fig:abstract_model}.

As in \cite{dutta2020capacity,skrimponis2020power,skrimponis2021towards,skrimponis2020efficient}, we model the transmissions in discrete-time 
frequency-domain and assume there are $N$
frequency bins.  We divide the bins into two groups:
\begin{itemize}
    \item $I_{\rm sig} \subseteq \{1,\ldots,N\}$ representing the frequency bins on which the desired signal is transmitted; and
    \item $I_{\rm int} = \{1,\ldots,N\} \backslash I_{\rm sig}$ representing the frequency bins on which the interference appears.
\end{itemize}
We assume that each transmitter performs digital beamforming and has 
only one output stream. We denote the frequency domain symbols for the
desired signal by $s_n$, $n\in I_{\rm sig}$
and the interfering signal samples by $v_n$,
$n \in I_{\rm int}$.  As a result, the input to the channel
is the signal,
\begin{equation}
    \nbu = \nbF \herm \nbx, \quad x_n = \begin{cases}
        s_n & \mbox{if } n \in I_{\rm sig} \\
        v_n & \mbox{if } n \in I_{\rm int}
        \end{cases},
\end{equation}
where $\nbF\herm \in \C^{n \times n}$ is the unitary 
IFFT matrix converting the frequency-domain vector
to time-domain. As for the power levels, we assume that the transmitted
symbols, $s_n$, 
are zero-mean complex Gaussian with 
energy per transmitted sample of $E_{\rm sig} = \E|s_n|^2$ and the interference symbols are also complex Gaussian with a frequency-dependent energy per sample, $E_{\rm int} = \E|v_n|^2$.
We will assume there is some reference
thermal noise level $N_0$ and let
\[
    \gamma_{\rm sig} := \frac{E_{\rm sig}}{N_0},
    \quad
    \gamma_{\rm int} := \frac{E_{\rm int}}{N_0},
\]
denote the signal and interference-to-noise ratios.

The samples $u_n$ are passed through a generic channel and receiver radio frequency front-end 
(RFFE) which are jointly modeled as a non-linear, memoryless function $\Phi(u_n, \nbd_n)$, where $\nbd_n$ are i.i.d.\ noise vectors with i.i.d. elements that include thermal noise
and noise from the non-linear components. Finally, the receiver performs digital beamforming using the vector $\nbw$
\begin{equation}
    r_n = \nbw\herm\Phi(u_n, \nbd_n).
\end{equation}
\begin{figure}[t]
    \centering
    \includegraphics[width = 0.99\linewidth]{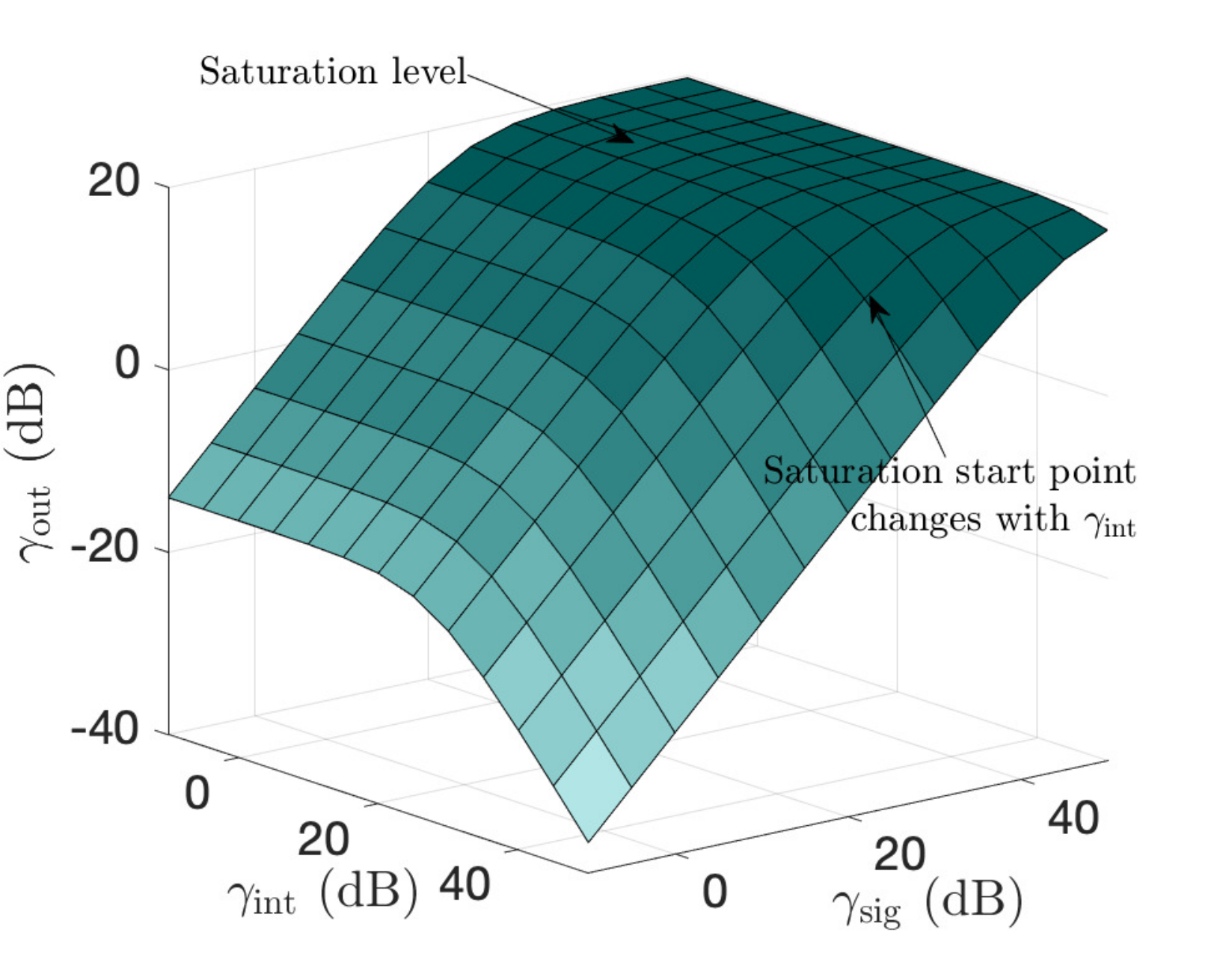}
    \caption{Input-output SNR relation.}
    \label{fig:snr_inout}
\end{figure}

\begin{figure*}[!t]
    \centering
    \begin{tikzpicture}[every text node part/.style={align=center, scale=1.0}]
    \footnotesize
    \ctikzset{bipoles/amp/width=0.9}
    
    \definecolor{matlabplot0}{rgb}{0.1154, 0.4519, 0.4522}
    \definecolor{matlabplot1}{rgb}{0.5313, 0.7875, 0.7877}
    
    \def\digcolor{matlabplot0}
    \def\anacolor{matlabplot1}

    \draw (0, 1) node[bareantenna, fill=gray!90]  (ant1) {};
    \draw node[label,above of=ant1,xshift=0cm]   () {$N_\mathrm{rx}$ antennas};
    
    \draw node[ampshape, below right of=ant1, scale=0.8, xshift=0.25cm, fill=\anacolor] (lna1) {};
    \draw (ant1) |- (lna1);
    \draw node[label,above of=lna1]   () {LNA};
    \draw node[label, below of=lna1,yshift=-0.5cm] () {\Large $\vdots$};
    
    \draw node[lowpassshape, right of=lna1, xshift=0.5cm, fill=\anacolor]
        (lpf1) {};
     \draw (lna1) |- (lpf1);
    \draw node[label,above of=lpf1] () {RF Filter};
        
    \draw node[mixer, right of=lpf1, xshift=0.5cm, scale=0.8, fill=\anacolor] (mixer1) {};
    \draw (lpf1) |- (mixer1.1);
    \draw node[label,above of=mixer1]   () {IF Mixer};
    
    \draw node[buffer, below of=mixer1,scale=0.5, xshift=1cm, yshift=-0.75cm, fill=\anacolor, rotate=180] (buf1) {};
    \draw (mixer1.2) |- (buf1.out);
    
    \draw node[buffer, below of=mixer1,scale=0.5, xshift=1cm, yshift=-0.75cm, fill=\anacolor, rotate=180] (buf1) {};
    
    \draw node[rectangle, draw, thick, right of=buf1] (mult1) {$\times\mathrm{M}$};
    \draw (buf1.in) |- (mult1);
    
    \draw node[oscillator,scale=0.8,right of=mult1, xshift=0cm, yshift=0cm, fill=\anacolor, rotate=180] (lo1) {};
    \draw (mult1) |- (lo1);
    
    \draw node[lowpassshape, right of=mixer1, xshift=0.5cm, fill=\anacolor]
        (lpf2) {};
     \draw (mixer1.3) |- (lpf2);
    \draw node[label,above of=lpf2] () {IF Filter};
    
     \draw node[vampshape, right of=lpf2, xshift=0.75cm, scale=0.8,
     fill=\anacolor] (agc1) {};
     \draw (lpf2) |- (agc1);
    \draw node[label,above of=agc1] () {AGC};
    
    \draw node[circ,right of=agc1,xshift=0.25cm]  (circ1) {};
    \draw (agc1) |- (circ1);
    
    \draw node[mixer, above right of=circ1, scale=0.8, xshift=0.25cm, yshift=0.75cm, fill=\anacolor] (mixer2) {};
    \draw (circ1) |- (mixer2.1);
    
    \draw node[rectangle, draw, thick, right of=circ1, xshift=-0.1cm, yshift=-0.45cm] (phase1) {$90^{\circ}$};
            
    \draw node[oscillator,scale=0.8,right of=circ1, xshift=1cm, yshift=0cm, fill=\anacolor, rotate=180] (lo2) {};
    
    \draw node[circ,right of=circ1,xshift=-0.1cm]  (circ2) {};
    
    \draw (mixer2.2) |- (circ2);
    \draw (phase1) |- (lo2);
    
    \draw  node[msport, right of=mixer2, circuitikz/RF/scale=1.75, scale=1.25, xshift=0.15cm, fill=\anacolor] (adc1) {ADC};
    \draw (mixer2.3) |- (adc1);
    
    \draw node[mixer, below right of=circ1, scale=0.8, xshift=0.25cm, yshift=-0.75cm, fill=\anacolor] (mixer3) {};
    \draw (circ1) |- (mixer3.1);
    \draw (phase1) |- (mixer3.4);
    
    \draw  node[msport, right of=mixer3, circuitikz/RF/scale=1.75, scale=1.25, xshift=0.15cm, fill=\anacolor] (adc2) {ADC};
    \draw (mixer3.3) |- (adc2);
    
    \draw [decorate,decoration={brace,amplitude=10pt,raise=0.7cm}]
        (mixer2.1) -- (adc1.east) node (brace) {};
    \draw node[label, above of=adc1, yshift=0.3cm, xshift=-0.1cm] () 
        {Direct conversion};
        
    \draw node[rectangle,draw,thick,right of=adc1, minimum width=1cm,minimum height=5cm, fill=\digcolor, xshift=1.9cm, yshift=-2cm] (12, 3) (dig_bf) {};
    
    \draw node[label,above of=dig_bf,yshift=2cm] () {Digital BF \\$N_\mathrm{rx}\times N_\mathrm{str}$};

    \foreach \y in {1.7,1.9,2.1} {
        \draw [-] ($(dig_bf.west)+(-1,-\y)$) -- 
            ($(dig_bf.west) +(0,-\y)$);
    }
    \draw [-] (adc1.right) -- ($(dig_bf.west) +(0,2)$);
    \draw [-] (adc2.right) -- ($(dig_bf.west) +(0,-0.61)$);
    
    \draw node[label,left of=dig_bf, xshift=-1.4cm,yshift=-1.95cm] () {From other\\ antennas};
        
    \draw node[rectangle,draw,thick,right of=dig_bf,
        minimum width=1cm,minimum height=1cm,
        fill=\digcolor, xshift=1 cm,yshift=0.1cm] 
            (fft1) {};
    \draw node[rectangle,draw,thick,right of=dig_bf,
        minimum width=1cm,minimum height=1cm,
        fill=\digcolor, xshift=0.9 cm] (fft) {};
    \draw [-,thick] (dig_bf.east) node[above,xshift=0.45cm] 
        {$N_\mathrm{str}$} -- (fft.west);
    \draw node[label,above of=fft,yshift=0.1cm] ()
        {OFDM};
    \draw [-,thick] (fft.east) node[above,xshift=0.55cm] 
        {$N_\mathrm{str}$} -- ($(fft.east)+(1,0)$);

    \draw node[bareantenna, fill=gray!90,below of=ant1,
        yshift=-2 cm]  (ant2) {};
    \draw node[ampshape, below right of=ant2, scale=0.8,
    xshift=0.25cm, fill=\anacolor] (lna2) {};
    \draw (ant2) |- (lna2);
    \draw node[label,right of=lna2,xshift=0.5cm] (cdots_lna) {\Large $\cdots$};
    \draw (lna2) -- (cdots_lna.west);
\end{tikzpicture}
    \caption{High-level architecture of a fully-digital superheterodyne 
    receiver architecture. The architecture supports $N_\mathrm{rx}$ antennas and
    $N_\mathrm{str}$ digital streams. The light green boxes represent analog and the 
    dark-green boxes the digital components. In the RF front-end, 
    some component are not shown.}
    \label{fig:arch}
\end{figure*}
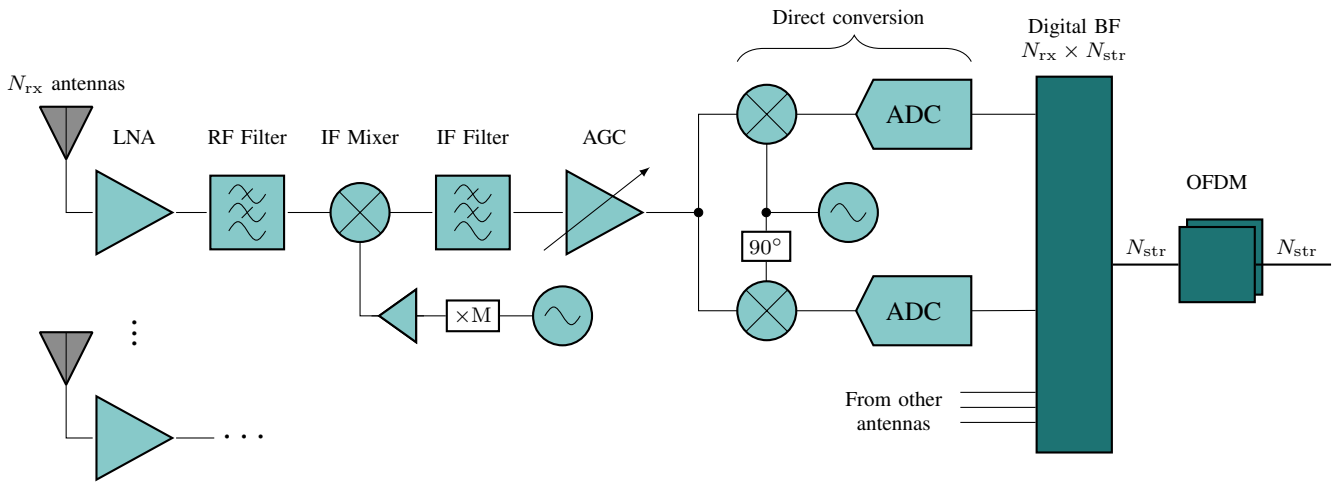

\section{Capacity Bound and Output SNR}
\label{sect:capacity}
Our goal is to characterize the performance of the discussed model in Sec.~\ref{sec:model} in terms of the spectral efficiency. To this end, similar to \cite{skrimponis2020efficient}, we make use of the concepts of the output SNR and input-output SNR relation described next. 

From Sec.~\ref{sec:model}, we have 
\begin{align}
    \label{eq:inoutre}
    \hat{\nbx} = \nbF\Phi(\nbF \herm \nbx, \nbD).
\end{align}

Assuming that the variables $\hat{x}_n$, $\nbd_n$, and $x_n$ have an underlying statistical model and are distributed as $\hat{X}$, $D$, and $X$, respectively, we can use the Bussgang-Rowe decomposition \cite{bussgang1952crosscorrelation,rowe1982memoryless} and model the non-linearity in the system (i.e., $\Phi(\cdot)$) as multiplying a scalar with its input and adding a noise which is uncorrelated with the input. More precisely, we can write
\begin{align}
\label{eq:lin_mod}
    \hat{X} = A X+ T,\quad \Exp |T|^2 &= \tau,
\end{align}
where
\begin{align}
\label{eq:alpha_tau_rx}
    A := \frac{\Exp\left[ \hat{X}^* X\right]}{\Exp\left|X\right|^2 }, \quad \tau :=  \Exp|\hat{X}- A X|^2,
\end{align} 
with $X^*$ denoting the complex conjugate of $X$.

In general, both $A$ and $\tau$ are functions of the 
input SNR $\gamma_\mathrm{in}$ 
so we may write $A = A(\gamma_{\mathrm{in}})$
and $\tau = \tau(\gamma_{\mathrm{in}})$. 
From \eqref{eq:lin_mod}, we can then define the 
{\it output SNR} off the desired signal $\nbs$ as,
\begin{equation} \label{eq:inout_snr}
    \gamma_{\mathrm{out}} = G(\gamma_{\mathrm{in}}) :=
    \frac{|A(\gamma_{\mathrm{in}})|^2}{\tau(\gamma_{\mathrm{in}})} 
    \gamma_{\rm sig}.
\end{equation}
This is the SNR that would be seen in attempting
to recover the input transmitted vector $\nbs$ from the
output vector $\hat{\nbs}$.  

Using the SNR enables us to bound the performance of the system in terms of the capacity. More precisely, using same steps as of \cite[Appendix~A]{skrimponis2020efficient}, we can show that the capacity of the system  can is lower bounded as,
\begin{equation}  \label{EQ:CAP_BAND_LIMITED1}
  C \geq  \frac{N_{\rm sig}}{N} f_s\log_2\left(1 +  \gamma_{\rm out}\right),
\end{equation}
where $f_s$ is the sample rate and $N_{\rm sig} = |I_{\rm sig}|$ represent the number of frequency bins for the signal. Moreover, assuming that the ADC performs oversampling with the ratio $\zeta$, using same steps as of \cite[Appendix~B]{skrimponis2020efficient}, we have

\begin{equation}  \label{EQ:CAP_BAND_LIMITED}
  C \geq  \frac{N_{\rm sig}\zeta}{N} f_s\log_2\left(1 +  \frac{\gamma_{\rm out}}{\zeta}\right),
\end{equation}

\begin{table}[t]
    \centering
    \caption{Parameters of the \SI{28}{GHz} RFFE devices used in the analysis.}

    \setlength{\tabcolsep}{3pt}
    \begin{tabular}{|>{\raggedright}m{0.9in}|c|c|c|c|}
    \hline
    
    \textbf{Parameter} &
    \textbf{LNA$^{\boldsymbol{(1)}}$} &
    \textbf{LNA$^{\boldsymbol{(2)}}$} &
    \textbf{Mixer$^{\boldsymbol{(1)}}$} &
    \textbf{Mixer$^{\boldsymbol{(2)}}$}
    \tabularnewline \hline
    Design [\textmu m] & 10 & 5 & 2 & 5
    \tabularnewline
    Noise Figure [dB] & 2.13 & 2.53 &  9.039 & 7.542
    \tabularnewline
    Gain [dB] & 14.26 & 12.85 &  0.16 &  3.558
    \tabularnewline
    IIP3 [dBm] & -1.456 & 0.603  & -3.1  & 2.1
    \tabularnewline
    Power [mW] &8.91 & 5.34 &   4.838 &   7.03
    \tabularnewline \hline
    \end{tabular}
    \label{tab:rffe28}
\end{table}

%

In this paper, we will show through detailed simulations that the input-output SNR relation can be approximated in the form of
\begin{equation} \label{eq:gamout}
    \hat{\gamma}_{\rm out} = \frac{\beta \gamma_{\rm sig}}{
    1 + \alpha_1 \gamma_{\rm sig} + \alpha_2\gamma_{\rm int}},
\end{equation}
    where $\gamma_{\rm int} = \frac{1}{|I_{\rm int}|}\sum_{n \in I_{\rm int}}E_i[n]$.
for three parameters $\beta$ and $\alpha_1$ and $\alpha_2$
using which we can evaluate the receiver front-end performance. Intuitively, this formula suggest that due to the non-linearity in the system: (i) the signal energy is reduced; (ii) a ratio of the signal is distorted; (iii) a ratio of the adjacent band signal (i.e., interference) is leaked to the desired band.

From \eqref{eq:gamout}, we also observe that the output SNR saturates to the value of $\frac{\beta}{\alpha_1}$ as the input signal SNR increases. Furthermore, for higher values of the interference signal the saturation should accrue for lower values of the input SNR. One can also observe these from Fig.~\ref{fig:snr_inout} which illustrates \eqref{eq:gamout} for a fixed values of $\beta$, $\alpha_1$ and $\alpha_2$ and different values of desired and interference signal powers.

\begin{table}[t]
    \centering
    \caption{Parameters of the \SI{140}{GHz} RFFE devices used in the analysis.}

    \setlength{\tabcolsep}{3pt}
    \begin{tabular}{|>{\raggedright}m{0.9in}|c|c|c|c|}
    \hline
    
    \textbf{Parameter} &
    \textbf{LNA$^{\boldsymbol{(1)}}$} &
    \textbf{LNA$^{\boldsymbol{(2)}}$} &
    \textbf{Mixer$^{\boldsymbol{(1)}}$} &
    \textbf{Mixer$^{\boldsymbol{(2)}}$}
    \tabularnewline \hline
    Design [\textmu m] & 4 &  2-4 & 1 & 1
    \tabularnewline
    Noise Figure [dB] & 7.50 & 7.48 & 21.53 & 20.47
    \tabularnewline
    Gain [dB] & 11.13 & 16.56 & -1.74 & -0.52
    \tabularnewline
    IIP3 [dBm] & -9.15 & -8.90 & -4.45 & -3.88
    \tabularnewline
    Power [mW] &  4.80 & 15.90 & 5.00 & 5.00
    \tabularnewline \hline
    \end{tabular}
    \label{tab:rffe140}
\end{table}

\section{Link-Layer Simulation} \label{sec:link}


The model in Section~\ref{sec:model} is a simplified
abstraction of an actual RFFE.  In this section, 
we validate the model and extract the parameters 
for \eqref{eq:gamout} with realistic, circuit simulations
of potential RFFEs at \SI{28}{GHz} and \SI{140}{GHz}.

\subsection{Signal and interference model}
We consider a downlink scenario in a communication link 
between an NR basestation (gNB) and a mobile device (UE). 
For each slot, the gNB generates a physical downlink shared
channel (PDSCH) that includes both information and control 
signals. The receiver uses the demodulation reference 
signal (DM-RS) for practical channel estimation. 
To compensate the common phase error (CPE) the 3GPP 5G NR,
standard introduce the phase tracking reference signal (PT-RS). 
The receiver performs coherent CPE estimation using the algorithm 
described in~\cite{syrjala2019phase}. For time synchronization,
between the gNB and UE we utilize the primary (PSS) and the secondary synchronization signals (SSS).

To model the interference, we assume the presence of another gNB
that generates i.i.d. $\CN(0, 1)$ symbols in frequency-domain. The
symbols are modulated with OFDM to generate interference in an
adjacent band. Even though the signals from the two gNBs are
originally independent in the frequency domain, the presence of 
the non-linear processing introduces distortion to the main signal
from the adjacent band. As explained in Section~\ref{sec:model},
this increase the total received energy and causes the RFFE 
saturation point to happen much earlier.

%
%
     
\begin{figure}[t]
    \centering
    \includegraphics[width = 0.99\linewidth]{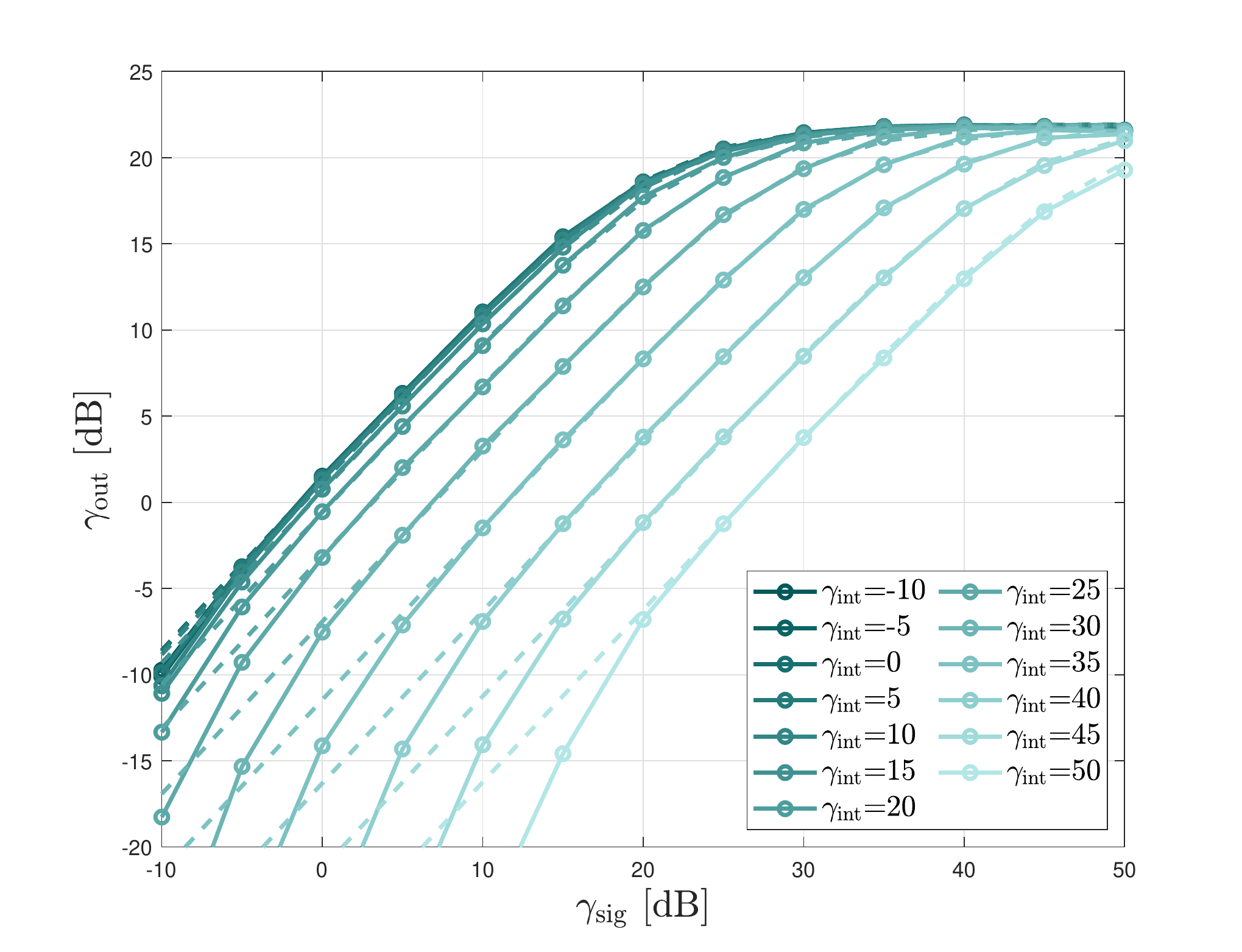}
    \caption{Evaluation of the approximation model in \eqref{eq:gamout} for Design$^{(1)}$ at \SI{28}{GHz} for different input interference power.}
    \label{fig:model_fit}
\end{figure}

\subsection{Receiver configurations}
Similarly to~\cite{skrimponis2021towards} we consider a fully-digital superheterodyne receiver architecture as shown in Fig.~\ref{fig:arch}. The receiver has $N_\mathrm{rx}=16$ or $64$ antennas with independent RFFE processing. The received signal is amplified with a low-noise amplifier  (LNA), downconverted with an intermediate frequency (IF) mixer, amplified with an automated gain control (AGC) amplifier before the direct conversion mixer, and finally quantized with a pair of ADCs. The system use filters in the RF and IF domain to improve the image rejection and the dynamic range of the system. The actions of the RFFE devices for each receiver configuration is modeled with a different non-linear function $\Phi(\cdot)$.

In~\cite{skrimponis2021towards} the authors design and evaluate RFFE devices at \SI{140}{GHZ} based on a \SI{90}{nm} SiGE BiCMOS HBT technology. They focus on minimizing the power consumption of the RFFE devices at a certain performance in terms of gain, noise figure (NF), and the input third order intercept point (IIP3). The LNA designs vary in terms of number of stages, topology, and transistor size. While the proposed double-balanced active mixers are all based on the conventional Gilbert-cell design, they vary in transistor size. The mixer performance characteristics depend on the input power from the local oscillator (LO). To further optimize the power consumption of the receiver, the authors propose a novel LO distribution model. Specifically, the mixers in the same tile share a common LO driver using power dividers and amplifiers. They provide a method to determine the best configuration of LO drivers that achieve the same performance for a minimum power.

Using the components and the optimization framework described in~\cite{skrimponis2021towards} we select two of the optimized designs for the \SI{140}{GHz} systems in our analysis. These two designs achieve similar performance to the state-of-the-art design discussed in \cite{skrimponis2020power} while achieving a significant improvement in power consumption. Similarly, for the \SI{28}{GHz} devices we design two common base emitter base collector (CBEBC) LNAs, and two active mixers based on the common Gilbert cell design. We use the optimization framework in~\cite{skrimponis2021towards} to determine the number of LO drivers. For both frequencies, Design$^{(1)}$ use 4-bit ADCs while Design$^{(2)}$ use 5-bit ADC pairs. This assumption is based on prior works \cite{abbas2017millimeter,zhang2018low,abdelghany2018towards,dutta2019case} indicating that 4 bits are sufficient for the majority of cellular data and control operations. Based on a flash-based 4-bit ADC in \cite{Nasri2017}, we consider the ADC $\mathrm{FOM} = 65\;\mathrm{fJ/conv}$. We summarize the parameters for the devices at \SI{28}{GHz} and \SI{140}{GHz} in Table~\ref{tab:rffe28} and Table~\ref{tab:rffe140} respectively. 


\begin{table}[t]
    \centering
    \caption{Model and system parameters for the receiver designs at \SI{28}{GHz} and \SI{140}{GHz}.}
    \label{tab:model_param}
    \begin{tabular}{|>{\raggedright}m{0.5in}|c|c|c|c|}
        \hline
         \multirow{2}{*}{\textbf{Parameter}} & \multicolumn{2}{c|}{\textbf{\SI{28}{GHz}}} & \multicolumn{2}{c|}{\textbf{\SI{140}{GHz}}}
        \tabularnewline \cline{2-5}
        & $\text{Design}^{(1)}$ & $\text{Design}^{(2)}$ & $\text{Design}^{(1)}$ & $\text{Design}^{(2)}$
        
        \tabularnewline \hline
        \multicolumn{1}{|c|}{$\beta$} & 1.3865 & 1.2725 & 0.3099 & 0.1862
        \tabularnewline 
        
        \multicolumn{1}{|c|}{$\alpha_1$} & 0.0090 & 0.0024 & 0.0021 & 0.0004
        \tabularnewline 
        
        \multicolumn{1}{|c|}{$\alpha_2$} & 0.0058 & 0.0017 & 0.0014 & 0.0003
        \tabularnewline \hline
        
        \multicolumn{1}{|c|}{RX antennas} & 16 & 16 & 64 & 64
        \tabularnewline
        
        \multicolumn{1}{|c|}{NF [dB]} & 2.78 & 3.08 & 9.40 & 11.50
        \tabularnewline
        
        \multicolumn{1}{|c|}{Power [mW]} & 411 & 404 & 1682 & 1355
        \tabularnewline \hline
    \end{tabular}
\end{table}

\subsection{Model fitting}
For each receiver design we fit a model in form of
\eqref{eq:gamout}. Since this nonlinear function 
$\Phi(\cdot)$ depends on the parameters $\alpha_1,\alpha_2$, and
$\beta$ we can write the estimated output-SNR model as $\hat{\gamma}_\mathrm{out}(\gamma_\mathrm{sig}, \gamma_\mathrm{int}; \alpha_1, \alpha_2, \beta)$. For the \emph{initial 
heuristic} fit we set, 
\begin{align}
    \beta = \frac{1}{F}, \quad \alpha_1 = \alpha_2 = \frac{1}{\gamma_\mathrm{sat}F},
    \label{eq:fitinit}
\end{align}
where $F$ is the noise factor of the system, and $\gamma_\mathrm{sat}$
the saturation SNR in linear scale. 
We then  optimize the fit  using the non-linear least squares regression method and optimize a problem of the following form,
\begin{align}
   Q(\gamma_\mathrm{sig}, \gamma_\mathrm{int}) := \min_{\alpha_1, \alpha_2, \beta}\|& \gamma_\mathrm{out}(\gamma_\mathrm{sig}, \gamma_\mathrm{int}) \nonumber \\ &
    - \hat{\gamma}_\mathrm{out}(\gamma_\mathrm{sig}, \gamma_\mathrm{int}; \alpha_1, \alpha_2, \beta)) \|_2^2, \nonumber\\
\end{align}
where  $\gamma_\mathrm{out}$ are the measurements from the link-layer simulation using \eqref{eq:inout_snr}, and $\hat{\gamma}_\mathrm{out}$ is the estimate using the model in
\eqref{eq:gamout}. The optimized parameters for the \SI{140}{GHz} and \SI{28}{GHz} receiver
designs are summarized in Tab.~\ref{tab:model_param}. In Fig.~\ref{fig:model_fit} we show that the model in~\eqref{eq:gamout} provides a very good fit. In particular, we see the linear regime for low-input SNR and the saturation for high-input signal power. We show that as the input interference increases the model the saturation SNR is also changing.

\section{Network-Level Simulation} \label{sec:network}
The above analysis shows that the performance of the RFFE
degrades in the presence of strong out-of-band
interference when the front-end dynamic range is low.
This fact raises a basic question:  \emph{how often
is the adjacent carrier interference strong in practical 
systems?}  In this section, we perform a simple network simulation
to assess the effect of adjacent carrier interference in 
a downlink cellular system with two 
adjacent carriers, carriers A and B.

We take account of a wrap-around \SI{1}{km}~$\times$~\SI{1}{km} network area to conduct the network-level simulation. Subject to a inter-site distance (ISD), gNBs are deployed by homogeneous Poisson point process (HPPP) with density $\lambda = \frac{4}{\pi\times\text{ISD}^2}$, and accordingly, the same number of UEs are uniformly distributed. Furthermore, all gNBs and UEs are 
randomly assigned to carrier A or B.
We use the notation $\text{gNB}_{A}$ and $\text{gNB}_{B}$ for base stations, and  $\text{UE}_{A}$ and $\text{UE}_{B}$ for UEs.
Individual gNBs are multi-sectorized with $8\times8$ uniform rectangular arrays (URA) per sector which is tilted down by $-12^\circ$, while each UE is equiped with a URA. The full 
parameter settings are shown in Table~\ref{tab:sim_param}. 

\begin{figure}[t]
    \centering
    \begin{tikzpicture}[every node/.append style={scale=1.0}]

    \definecolor{matlabplot0}{rgb}{0.1154, 0.4519, 0.4522}
    \definecolor{matlabplot1}{rgb}{0.5313, 0.7875, 0.7877}
    
    \tikzset{BS style/.style={
        draw,
        minimum width=1cm,
        minimum height=4cm,
        fill=gray,
        inner sep=0,
        outer sep=0,
        scale=0.75
    }}
    \tikzset{UE Style/.style={
        draw,
        fill=gray,
        minimum width=1.2cm,
        minimum height=2cm,
        scale=0.5
    }}

    \path (1.7, 0) coordinate (UE_A_pos);
    \path (4.3, 0) coordinate (UE_B_pos);
    \path (0, 3) coordinate (BS_A_pos);
    \path (6, 3) coordinate (BS_B_pos);
    \draw (BS_A_pos) node[BS style,BaseStationSimple] (BS_A) {};
    \draw (BS_B_pos) node[BS style,BaseStationSimple] (BS_B) {};
    \draw (UE_A_pos) node[UserTerminal,UE Style, button fill color=gray!40] (UE_A) {};
    \draw (UE_B_pos) node[UserTerminal,UE Style, button fill color=gray!40] (UE_B) {};

    \def\pos{-65.5}
    
    \draw (BS_A.antenna center) node[Beam,minimum width=0.5cm,minimum height=2mm,rotate=\pos-55,fill=matlabplot0!30] () {};
    \draw (BS_A.antenna center) node[Beam,minimum width=0.5cm,minimum height=2mm,rotate=\pos+55,fill=matlabplot0!30] () {};
    \draw (BS_A.antenna center) node[Beam,minimum width=0.75cm,minimum height=2.5mm,rotate=\pos-25,fill=matlabplot0!60] () {};
    \draw (BS_A.antenna center) node[Beam,minimum width=0.75cm,minimum height=2.5mm,rotate=\pos+25,fill=matlabplot0!60] () {};
    \draw (BS_A.antenna center) node[Beam,minimum width=1.5cm,minimum height=3mm,rotate=\pos,fill=matlabplot0!80] () {};
  
    \definecolor{gnbB_color}{rgb}{0.82, 0.1, 0.26}
    \def\pos{-114.5}
    \draw (BS_B.antenna center) node[Beam,minimum width=0.5cm,minimum height=2mm,rotate=\pos-55,fill=gnbB_color!30] () {};
    \draw (BS_B.antenna center) node[Beam,minimum width=0.5cm,minimum height=2mm,rotate=\pos+55,fill=gnbB_color!30] () {};
    \draw (BS_B.antenna center) node[Beam,minimum width=0.75cm,minimum height=2.5mm,rotate=\pos-25,fill=gnbB_color!60] () {};
    \draw (BS_B.antenna center) node[Beam,minimum width=0.75cm,minimum height=2.5mm,rotate=\pos+25,fill=gnbB_color!60] () {};
    \draw (BS_B.antenna center) node[Beam,minimum width=1.5cm,minimum height=3mm,rotate=\pos,fill=gnbB_color!80] (beam_1) {};
    
    \draw node[label,below of=UE_A, xshift=0.1cm, yshift=0.1cm] () {${\rm UE}_{\rm A}$};
    \draw node[label,below of=UE_B, xshift=0.1cm, yshift=0.1cm] () {${\rm UE}_{\rm B}$};
    \draw node[label,below of=BS_A, xshift=0cm, yshift=-0.9cm] () {${\rm gNB}_{\rm A}$};
    \draw node[label,below of=BS_B, xshift=0cm, yshift=-0.9cm] () {${\rm gNB}_{\rm B}$};
    
    \draw[densely dotted,decoration={markings,mark= at position 0.5 with {\arrow{latex}},}, postaction={decorate}] (BS_A.antenna center) -- (UE_B.north);
    \draw[densely dotted,decoration={markings,mark= at position 0.5 with {\arrow{latex}},},postaction={decorate}] (BS_B.antenna center) -- (UE_A.north);
    \draw[decoration={markings,mark= at position 0.75 with {\arrow{latex}},},postaction={decorate}] (BS_A.antenna center) -- (UE_A.north);
    \draw[decoration={markings,mark= at position 0.75 with {\arrow{latex}},},postaction={decorate}] (BS_B.antenna center) -- (UE_B.north);
    
    \path (2.2, 4.3) coordinate (Legend);
    \draw (Legend) -- ++(-0.25,0);
    \draw (Legend) ++ (0.6,0) node[label] () {Signal};
    \draw[densely dotted] (Legend) ++(0,-0.4) -- ++(-0.25,0);
    \draw (Legend) ++ (1,-0.4) node[label] () {Interference};
\end{tikzpicture}
    \caption{Signal and interfering
    paths in a system with two carriers.}
    \label{fig:network}
\end{figure}
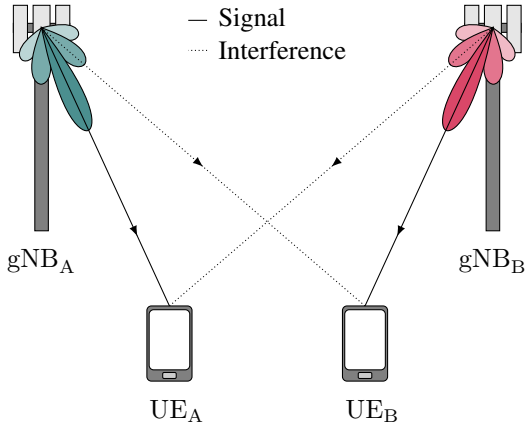

Fig.~\ref{fig:network} shows the scenario for analyzing the interference between adjacent carrier frequencies. A UE in carrier A 
will receive the desired signal from its serving BS and interference signal from non-serving
BSs in
carrier A in the same carrier and
from all BSs in carrier B.

\begin{figure*}[t]
\centering
    \subfigure[\SI{28}{GHz}]{\includegraphics[width=0.49\linewidth]{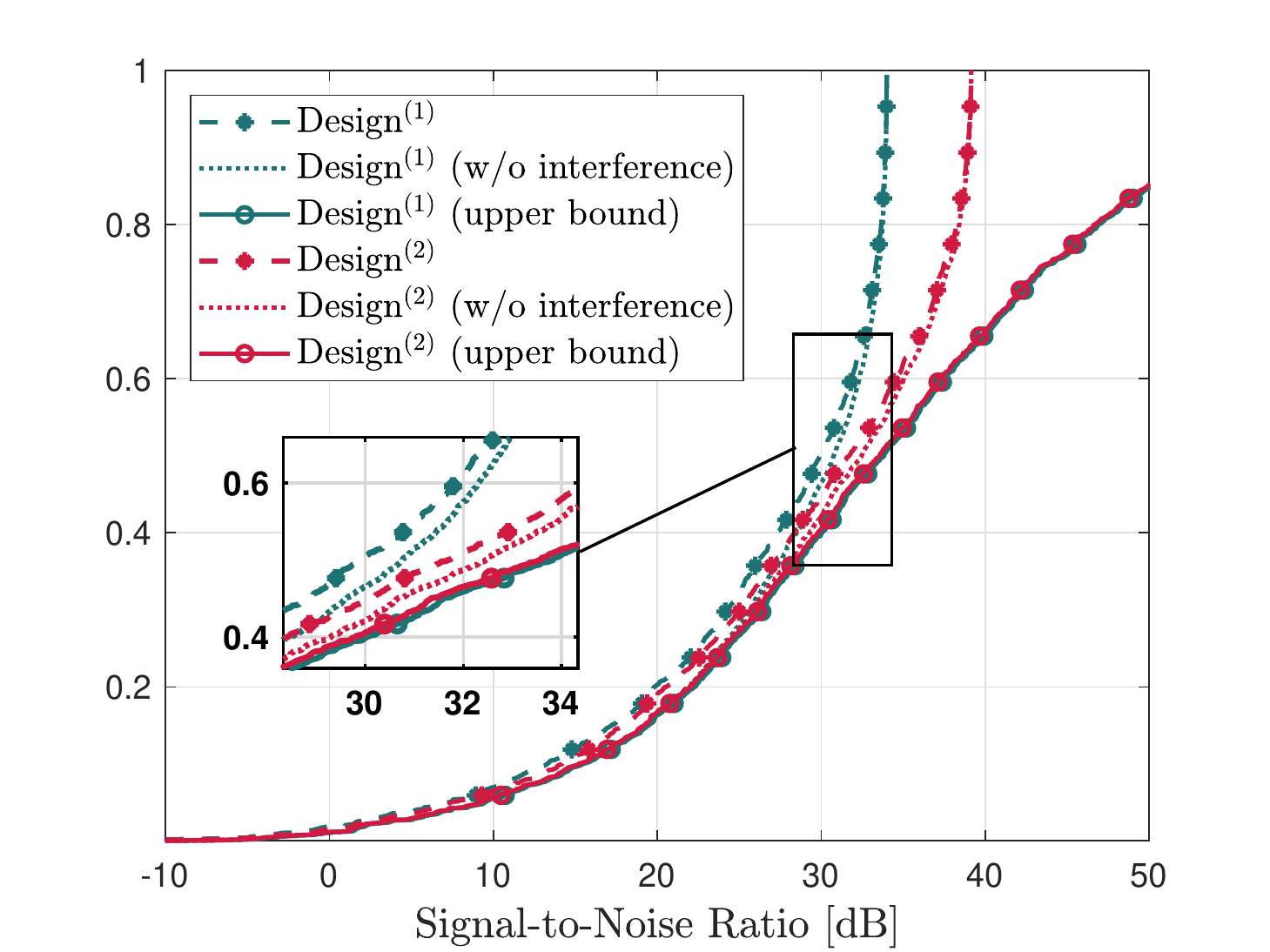}}
    \subfigure[\SI{140}{GHz}]{\includegraphics[width=0.49\linewidth]{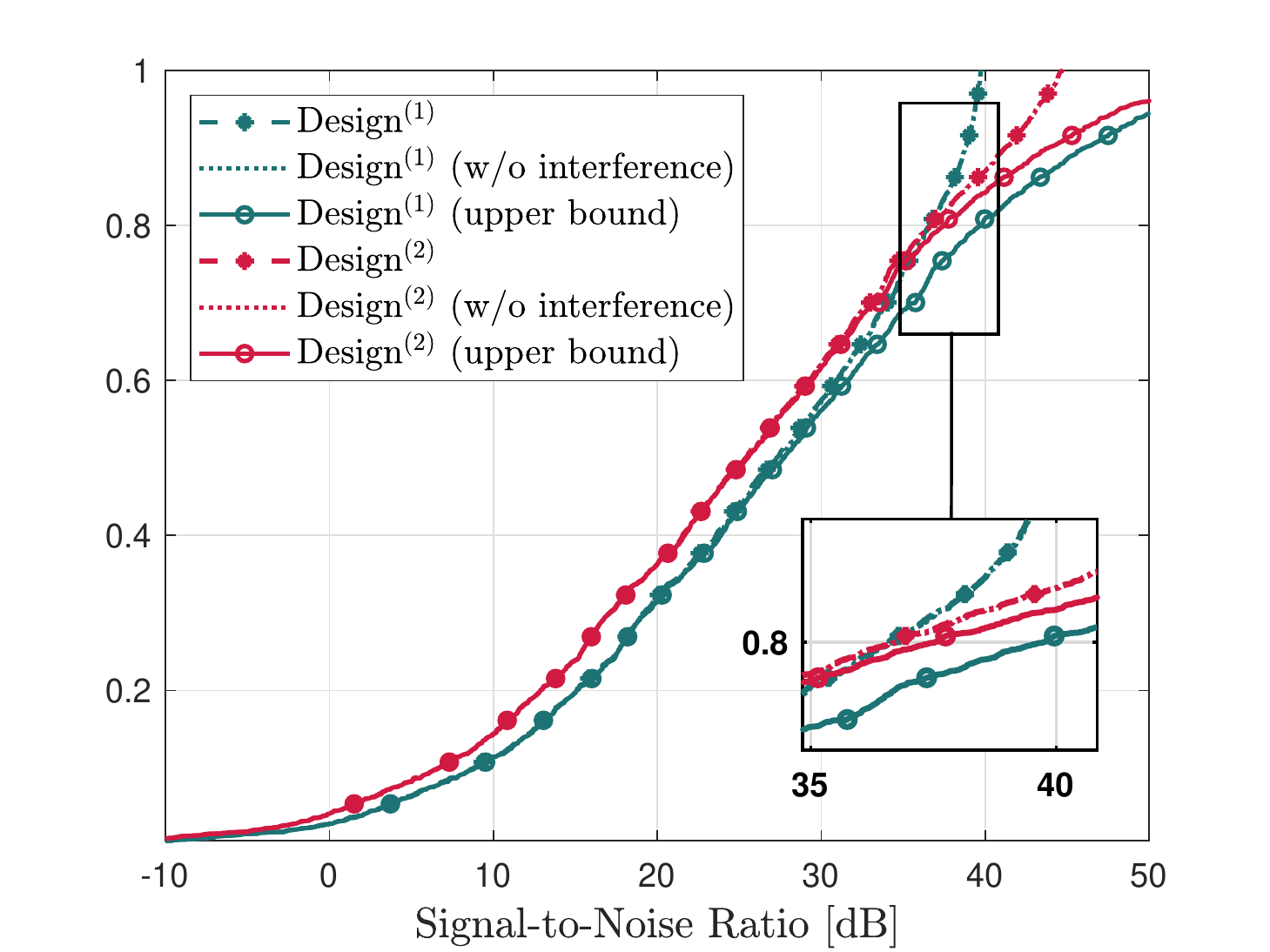}}
    
 \caption{Estimated distribution of downlink SNRs with different carrier frequencies
    and different RFFE designs.  The plots
    show the SNR under the full model
    \eqref{eq:SNR} with distortion from
    adjacent carrier interference and 
    in-band signal; the SNR 
    with no adjacent carrier interference distortion ($\alpha_2=0)$; and
    the SNR with no in-band or
    adjacent carrier interference distortion ($\alpha_1=\alpha_2=0)$.}
  \label{fig:snr_distributions}
\end{figure*}


For every gNB-UE pair, we generate a multi-path channel for two different frequency cases, \SI{28}{GHz} and \SI{140}{GHz},  according to~\cite{3GPP38.901} and compute the SNR for downlink case. 
Specifically, we employ the path-loss model specified in~\cite{3GPP38.901} for the Urban Micro Street Canyon (Umi-Street-Canyon) environment. The 3GPP NR standard is very flexible and we assume the channel models described in
\cite{3GPP38.901} will hold for the \SI{140}{GHz} communication systems.

Within each carrier, we then assume that each UE
is served by the strongest gNB.  As a simplification, 
the gNBs and UEs then beamform
along the strongest path with no regard to
interference nulling to other UEs.  
A sufficient number of UEs are dropped such
that we can obtain one UE served by each
sector in each gNB.  Hence, the simulation
drop represents one point in time where
each gNB is using its entire bandwidth 
on one UE.

With the channels and beamforming
direction, we can then estimate the effective
SINR at each UE.  
Following the model \eqref{eq:gamout},
we estimate the SINR as:
\begin{equation}
\label{eq:SNR}
    \gamma = \frac{\beta E_\mathrm{sig}^{a}}{E_\mathrm{kT} +
    \alpha_{1}E_\mathrm{tot}^{a} + \alpha_{2}E_\mathrm{tot}^{b}}.
\end{equation}
Here, $E_\mathrm{sig}^{a}$ 
and $E_\mathrm{int}^a$ 
is the energy 
per sample of the serving and interfering
signals including the beamforming and
element gains at the TX and RX,
and $E_\mathrm{kT}$ is the thermal noise.
The distortion from the non-linearities
is modeled by two terms:
$\alpha_{1}E_\mathrm{tot}^{a}$ captures
the distortion from the total power from all base stations in carrier A (serving and non-serving); $\alpha_{2}E_\mathrm{tot}^{b}$
captures the distortion from
the total power from all base stations
in the adjacent carrier, carrier B.
For these terms, we assume that the distortion
is spatially white so we do not 
add the RX beamforming gains on each path.
The terms however do include the TX element
and beamforming gains as well as the RX element
gain.

\begin{table}[t]
    \centering
    \caption{Network simulation parameters.}
    \setlength{\tabcolsep}{3pt}
    \label{tab:sim_param}
    \begin{tabular}{|>{\raggedright}m{1.6in}|c|c|}
        \hline
        \textbf{Parameter} & \multicolumn{2}{c|}{\textbf{Value}}
        \tabularnewline \hline
      
        Carrier frequency, [\SI{}{\GHz}] & $28$ & $140$
        \tabularnewline
        
        Total bandwidth, [\SI{}{\MHz}] & $190.80$ & $380.16$
        \tabularnewline 
        
        Sample rate, 
        [\SI{}{\MHz}] & $491.52$ & $1966.08$
        \tabularnewline 
        
        gNB antenna configuration &{$8\times8$} & {$16\times16$}
        \tabularnewline
        UE antenna configuration &{$4\times4$}& {$8\times8$}
        \tabularnewline \cline{2-3}
        
        Area [\SI{}{\meter}$^2$]  &  \multicolumn{2}{c|}{$1000 \times 1000 $}
        \tabularnewline
        
        UE and gNB min. distance [\SI{}{\meter}] &  \multicolumn{2}{c|}{$10$}
        \tabularnewline
        
        ISD [\SI{}{\meter}] & \multicolumn{2}{c|}{$200$}
        \tabularnewline
        
        gNB height [\SI{}{\meter}] &  \multicolumn{2}{c|}{$\ncalU(2,5)$}
        \tabularnewline
        
        UE height [\SI{}{\meter}] &  \multicolumn{2}{c|}{$1.6$}
        \tabularnewline
        
        gNB TX power [\SI{}{dBm}] & \multicolumn{2}{c|}{$30$}
        \tabularnewline
        
        gNB downtilt angle & \multicolumn{2}{c|}{$-12^\circ$}
        \tabularnewline
        
        gNB number of sectors & \multicolumn{2}{c|}{$3$}
        \tabularnewline
        
        Vertical half-power beamwidth 
        & \multicolumn{2}{c|}{$65^\circ$}
        \tabularnewline
        
        Horizontal half-power beamwidth 
        & \multicolumn{2}{c|}{$65^\circ$}
        \tabularnewline \hline
    \end{tabular}
\end{table}


Fig.~\ref{fig:snr_distributions} shows the SNR distributions  for the designs discussed in Section~\ref{sec:link} at \SI{28}{GHz} and \SI{140}{GHz}. As a performance benchmark,
we compare the SNR distribution under three models:
\begin{itemize}
    \item SNR with adjacent carrier interference and in-band distortion:  This is the model
    \eqref{eq:SNR} with the parameters
    for $\alpha_1$ and $\alpha_2$ in
    Table~\ref{tab:model_param} found from
    the circuit simulations.  The
    resulting SNR CDF is shown
    in the dashed line in Fig.~\ref{fig:snr_distributions}.
    
    \item SNR with no adjacent carrier interference and in-band distortion:  
    This is the model
    \eqref{eq:SNR} with the parameters
    for $\alpha_1$ in
    Table~\ref{tab:model_param} but
    $\alpha_2 = 0$.  The SNR CDF is shown in the
    dotted line in Fig.~\ref{fig:snr_distributions}.
    
    \item SNR with no adjacent carrier interference and  no in-band distortion:  
    This is the model
    \eqref{eq:SNR} with the parameters
    for $\alpha_1=\alpha_2=0$.  
    The resulting SNR CDF is shown in the
    solid line in Fig.~\ref{fig:snr_distributions}.

\end{itemize}

Comparing the plots, we observe that the impact of distortion from adjacent carrier interference  is negligible.   This suggests that for these parameters,
there may be no need for extra filtering in the RF/IF or baseband to suppress the adjacent carrier interference.   Thus, we can conclude that by optimizing the RFFE devices and reducing the dynamic range of the system, we can improve the energy efficiency without being vulnerable from the adjacent carrier interference. Furthermore, as expected the designs at \SI{28}{GHz} have lower saturation points comparing to the \SI{140}{GHz} designs, 
due to the difference in the beamforming gain resulted from the difference in the number of antennas.

As explained in Section~\ref{sec:link}, at \SI{140}{GHz} we expect the designs to have different performance. In the low-SNR regime Design$^{(1)}$ performs better due to the lower NF, while Design$^{(2)}$ performs better in the high-SNR regime due to the larger number of ADC bits.

\section{Conclusions} \label{sec:conclusion}

Low-power designs for receivers above \SI{100}{GHz}
have traditionally relied on operating the
circuits with limited dynamic range either
via low-resolution ADCs or low-power mixers.
While there techniques have been successful
when considering in-band signal distortion,
the limited dynamic range can be an issue for
adjacent carrier interference.  
We have developed a simple 
mathematically model to describe this effect 
and fit the model parameters on realistic 
circuit designs at \SI{28}{GHz} and \SI{140}{GHz}.
The models were then be used in a simple
network simulation to estimate the effect of
adjacent carrier interfence in cellular systems
with two operators.  Our preliminary simulations
suggest that, at least under the parameters
considered, highly optimized power designs
are not significantly vulnerable to adjacent
carrier interference.  Future work can consider
other deployments where the adjacent carrier
interference could be higher.  For example,
short range local area signals operating adjacent
to cellular bands as well as mixed applications
such as terrestrial networks sharing spectrum 
with vehicular or UAV systems.  

\bibliographystyle{IEEEtran}
\bibliography{bibl}
\end{document}